\documentclass[twocolumn,showpacs,preprintnumbers,amsmath,amssymb]{revtex4}
\usepackage{graphicx}
\usepackage{dcolumn}
\usepackage{bm}

\newcommand{\bef}{\begin{figure}}
\newcommand{\eef}{\end{figure}}

\newcommand{\be}{\begin{equation}}
\newcommand{\ee}{\end{equation}}
\newcommand{\bea}{\begin{eqnarray}}
\newcommand{\eea}{\end{eqnarray}}
\begin{document}

\title{Longitudinal scaling of observables in heavy-ion collision models}

\author{Md. Nasim$^1$, Chitrasen Jena$^2$, Lokesh Kumar$^3$, Pawan Kumar Netrakanti$^4$ and Bedangadas Mohanty$^1$}
\affiliation{$^1$Variable Energy Cyclotron Centre, Kolkata 700064, India, $^2$Institute of Physics, Bhubaneswar 751001, India, $^3$Kent State University, Kent, Ohio 44242, USA, and $^4$Bhabha Atomic Research Centre, Mumbai 400 085, India}

\date{\today}
\begin{abstract}

Longitudinal scaling of pseudorapidity distribution of charged particles  
($dN_{\mathrm {ch}}/d\eta$) is observed when presented as a function of 
pseudorapidity ($\eta$) shifted by the beam rapidity ($\eta$ - $y_{\mathrm {beam}}$) 
for a wide range of collision systems ($e^{+}+e^{-}$, $p$+$p$, $d$+A and A+A) 
and beam energies. Such a scaling is also observed for the elliptic flow 
($v_{2}$) of charged hadrons in A+A collisions. This is a striking observation, 
as $v_{2}$ is expected to be sensitive to the initial conditions, the expansion 
dynamics and the degrees of freedom of the system, all of which potentially varies
with collision system and colliding energies. We present a study of the longitudinal 
scalings of $dN_{\mathrm {ch}}/d\eta$, average transverse momentum ($\langle p_{\mathrm T} \rangle$) 
and $v_{2}$ using transport models UrQMD and AMPT for 
Au+Au collisions at center of mass energies ($\sqrt{s_{\mathrm {NN}}}$) of 19.6, 62.4, 200 GeV
and Pb+Pb collisions at 2760 GeV. Only the AMPT models which includes partonic effects and 
quark coalescence as a mechanism of hadronization, shows longitudinal scaling 
for $dN_{\mathrm {ch}}/d\eta$, $\langle p_{\mathrm T} \rangle$ 
and $v_{2}$. Whereas the UrQMD and AMPT default versions show longitudinal scaling only 
for $dN_{\mathrm {ch}}/d\eta$ and $\langle p_{\mathrm T} \rangle$. We also discuss 
the possibility of longitudinal 
scaling of $v_{2}$ within two extreme scenarios of models with hydrodynamic and 
collisionless limits. We find the longitudinal scaling of bulk observables to be an 
important test for the underlying physics mechanism in models of particle production. 
\end{abstract}
\maketitle

\section{INTRODUCTION}
Scale invariance in experimental observables from heavy-ion collisions at the Relativistic
Heavy Ion Collider have provided interesting insights about particle production
mechanism in these reactions~\cite{whitepapers}. 
Observables like $\langle p_{\mathrm T} \rangle$, 
freeze-out parameters~\cite{starprc}, 
pion interferometry radii~\cite{starhbt} are observed to scale with some powers of  $dN_{\mathrm {ch}}/d\eta$. 
Within the frame work of thermal models some of these reflect a constant energy per 
particle at freeze-out. The elliptic flow for identified baryons and mesons when 
divided by the number of constituent quarks is found to scale with the kinetic energy
of the particles (or constituents)~\cite{starncq}. This has been interpreted 
as due to development of substantial collectivity in the partonic phase~\cite{starphiflow} 
of the evolution of the heavy-ion collisions and coalescence being the dominant mechanism 
of particle production for the intermediate $p_{\mathrm T}$ range of 2 - 6 GeV/$c$~\cite{voloshin}. 
Scalings have been observed in $p$+$p$ collisions. At low $p_{\mathrm T}$ ($<$ 2 GeV/$c$) 
for a given $\sqrt{s}$, the invariant yields of identified hadrons are observed to scale 
when plotted as a function of $m_{\mathrm T}$ ($m_{\mathrm T} = \sqrt{p_{\mathrm T}^{2} - m^{2}}$ and
$m$ is the mass of the hadron),~\cite{stardau}. This indicates the absence 
of significant radial flow in $p$+$p$ collisions. At high $p_{T}$ ($>$ 2 GeV/$c$) the 
product of the invariant yields of a hadron and some power of the $\sqrt{s}$ become 
independent of $\sqrt{s}$ when plotted as a function of $x_{\mathrm T}$ 
(= 2$p_{\mathrm T}/\sqrt{s}$)~\cite{starxt}. This is interpreted as due to dominance 
of the pQCD process (jets) in $p$+$p$ collisions. Scaling has been also observed in the 
longitudinal direction, represented by a variable $\eta$ - $y_{\mathrm {beam}}$, 
in $dN_{\mathrm {ch}}/d\eta$~\cite{lf}, $v_{2}$~\cite{v2lf} and directed flow 
($v_{1}$)~\cite{v1lf} in A+A collisions.  Further such scaling have been widely used to predict the 
values of the observables at higher energy regimes in nucleus-nucleus 
collisions~\cite{lfpredict}. Recent results on $dN_{\mathrm {ch}}/d\eta$ at midrapidity
for Pb+Pb central collisions at $\sqrt{s_{NN}}$ = 2760 GeV is found to be higher than expected from 
extrapolation based on the longitudinal scaling of $dN_{\mathrm {ch}}/d\eta$ at lower beam energies~\cite{alicepaper}.

The longitudinal scaling of $dN_{\mathrm {ch}}/d\eta$ is a widely discussed subject as it
is observed for variety of colliding systems starting from $e^{+}+e^{-}$, $p$+$p$, $d$+A 
to  A+A collisions~\cite{lf,lfda}. This phenomena is often called limiting fragmentation. 
It was hypothesized by Benecke et al.~\cite{benecke}, Feynman~\cite{feynman} 
and Hagedron~\cite{hagedron} that as $\sqrt{s} \rightarrow \infty$ the multiplicity
distribution becomes independent of $\sqrt{s}$. From a microscopic picture the longitudinal 
scaling is understood. assuming  the rapidity distributions of the produced particles are functions 
of $x$ (fraction of the hadron longitudinal momentum carried by a typical parton) alone, 
and not of the total energy. If the hadron interactions are short-ranged in rapidity then
the hadron rapidity distributions would reproduce the corresponding distributions of the 
liberated partons. The picture is very similar to Bjorken scaling of parton distributions. 
This interpretation can be easily linked to initial state gluon dynamics 
of the system~\cite{cgc}. From a macroscopic picture, the entropy conservation in heavy-ion 
collisions can make $dN_{\mathrm {ch}}/d\eta$ insensitive to some aspects of dynamics 
of system and hence may be the cause of the scaling.

The $p_{\mathrm T}$ integrated $v_{2}$ for a given rapidity range is defined as~\cite{flow}
\begin{equation}
v_{2}=\langle\cos(2(\phi-\Psi))\rangle,
\end{equation}
where $\phi$ and $\Psi$ are the charged particle azimuthal angle and reaction plane angle respectively.
The observed $v_{2}$ is affected by the initial conditions, it is sensitive to the equation of state 
and its magnitude is determined by the interactions of the constituents through out the evolution 
of the system  in heavy-ion collisions~\cite{flow1,torrieri}. 
Hence the physical interpretation of longitudinal scaling of $v_{2}$ is counter intuitive. 
Longitudinal scaling of $v_{2}$ exhibited for a wide beam energy range, different collision 
centrality and collision species would tend to indicate weak dependence of $v_{2}$ on the above 
mentioned physical conditions. 
Recent studies~\cite{torrieri} suggest that the simultaneous observation of longitudinal 
scaling of $v_{2}$ and $dN_{\mathrm {ch}}/d\eta$ can be reconciled only if the system 
formed in heavy-ion collisions are weakly coupled. This is 
contrary to other indirect estimations of the shear viscosity to entropy ratio which 
suggests the system is strongly coupled~\cite{etabys}.

In this paper, we study the longitudinal scaling of $dN_{\mathrm {ch}}/d\eta$,$\langle p_{\mathrm T} \rangle$  
and $v_{2}$ using models AMPT (A Multi Phase Transport Model, ver 1.11)~\cite{ampt} 
and  UrQMD (Ultra Relativistic Quantum Molecular Dynamics, ver 2.3)~\cite{urqmd} 
for charged particles in Au+Au collisions at $\sqrt{s_{\mathrm {NN}}}$ = 19.6, 62.4, 200 GeV
and Pb+Pb collisions at 2760 GeV. 
The aim being to see if these models also exhibit such longitudinal scalings and hence 
provide a physical insight behind the phenomena. The UrQMD model is based on a microscopic 
transport theory where phase space description of the reactions and hadron-hadron 
interactions are important. The AMPT model can be studied in two configurations, in the 
AMPT default version the minijet partons are made to undergo scattering before they 
are allowed to fragment into  hadrons, while in the AMPT-SM string melting 
scenario additional scattering occurs among the quarks and the hadronization
occurs through the mechanism of parton coalescence.

\section{LONGITUDINAL SCALING OF $dN_{\mathrm {ch}}/d\eta$ and $\langle p_{\mathrm T} \rangle$}
\bef
\begin{center}
\includegraphics[scale=0.4]{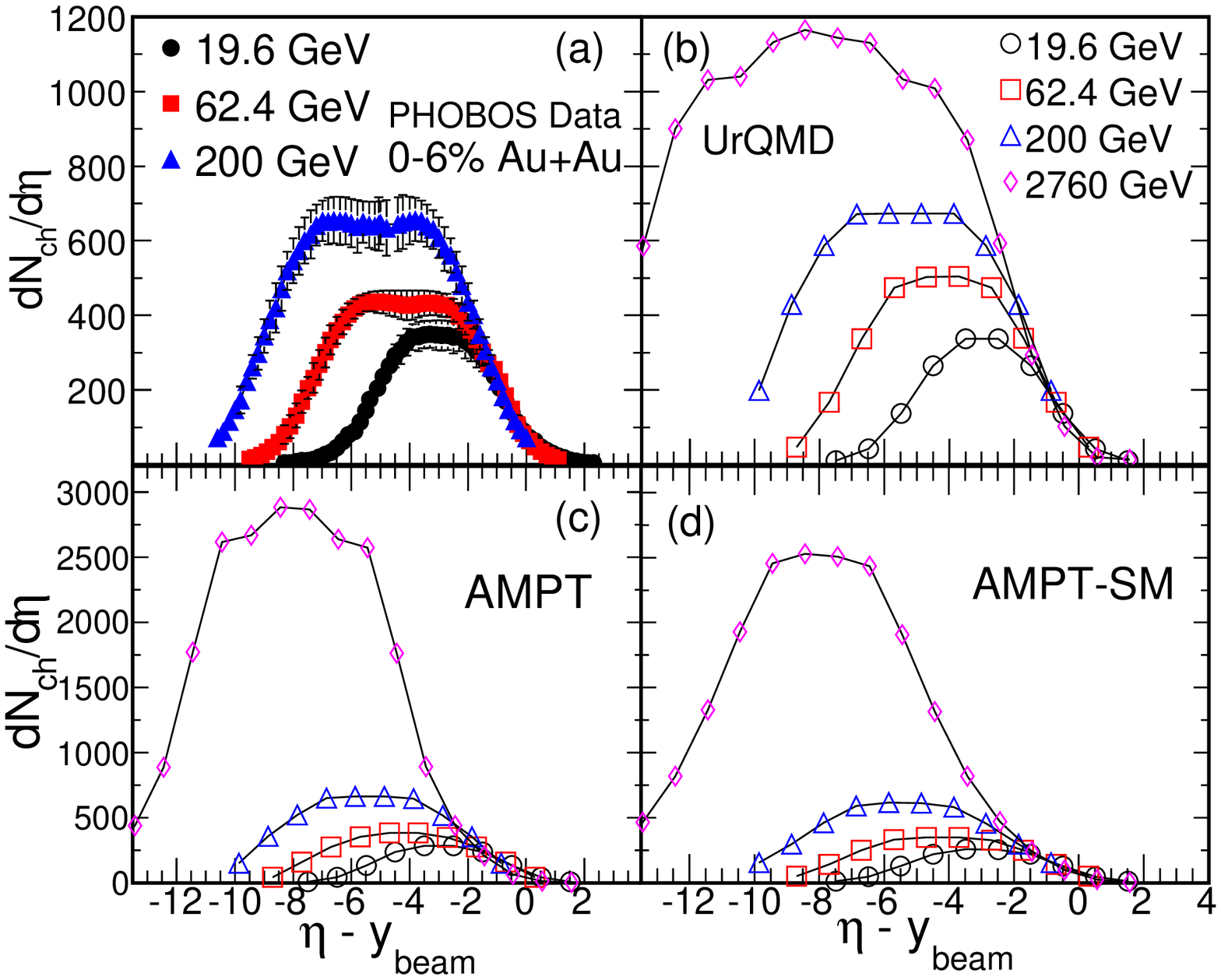}
\caption{(Color online) $dN_{\mathrm {ch}}/d\eta$ versus $\eta$-$y_{\mathrm {beam}}$
for 0--6\% central Au+Au collisions at $\sqrt{s_{\mathrm {NN}}}$ = 19.6, 62.4 and 200 GeV
from (a) the PHOBOS experiment at RHIC~\cite{lf}, (b) UrQMD, (c) AMPT default 
and (d) AMPT-SM. Also shown are the model results from Pb+Pb collisions 
at $\sqrt{s_{\mathrm {NN}}}$ = 2760 GeV.}
\label{fig1}
\end{center}
\eef

Figure~\ref{fig1} shows the $dN_{\mathrm {ch}}/d\eta$ versus $\eta$-$y_{\mathrm {beam}}$ 
for 0--6\% central Au+Au collisions at $\sqrt{s_{\mathrm {NN}}}$ = 19.6, 62.4 and 200 GeV
from (a) the PHOBOS experiment at RHIC~\cite{lf}, (b) UrQMD, (c) AMPT and (d) AMPT-SM models. Also shown
are the results from the models for Pb+Pb collisions at $\sqrt{s_{\mathrm {NN}}}$ = 2760 GeV.
The $y_{\mathrm {beam}}$ values for $\sqrt{s_{\mathrm {NN}}}$ = 19.6, 62.4, 200 and 2760 GeV 
are 3.03, 4.19, 5.36 and 7.98 respectively. The longitudinal scaling observed in 
$dN_{\mathrm {ch}}/d\eta$ in the data  (Fig.~\ref{fig1}(a)) is also observed in all the 
models studied. Such scalings have also been observed for other models like HIJING~\cite{hijing}
and HIJING B$\bar{B}$~\cite{hijingbb, pawan}. 

\bef
\begin{center}
\includegraphics[scale=0.4]{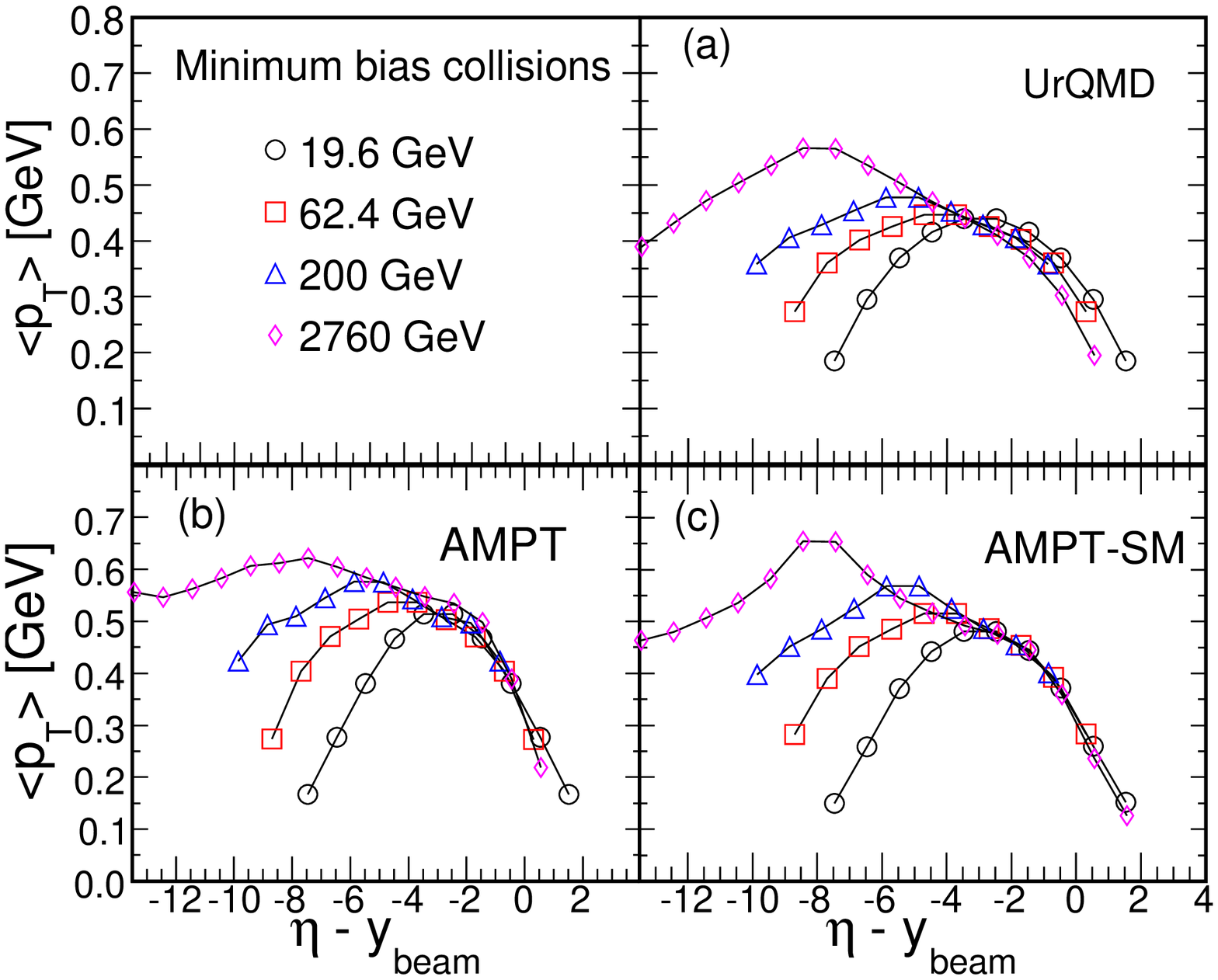}
\caption{(Color online) $\langle p_{\mathrm T} \rangle$ versus $\eta$-$y_{\mathrm {beam}}$
for minimum bias Au+Au collisions at $\sqrt{s_{\mathrm {NN}}}$ = 19.6, 62.4 and 200 GeV
from (a) the UrQMD, (b) AMPT default and (c) AMPT-SM. Also shown are the model results from Pb+Pb collisions 
at $\sqrt{s_{\mathrm {NN}}}$ = 2760 GeV.}
\label{fig2}
\end{center}
\eef

Figure~\ref{fig2} shows the $\langle p_{\mathrm T} \rangle$ for the charged particles 
versus $\eta$-$y_{\mathrm {beam}}$ 
for minimum bias Au+Au collisions at $\sqrt{s_{\mathrm {NN}}}$ = 19.6, 62.4 and 200 GeV
from (a) the UrQMD, (b) AMPT and (c) AMPT-SM models. Also shown are the results from the models 
for minimum bias Pb+Pb collisions at $\sqrt{s_{\mathrm {NN}}}$ = 2760 GeV. 
There are no experimental data available at RHIC for
$\langle p_{\mathrm T} \rangle$ versus $\eta$-$y_{\mathrm {beam}}$ hence not shown in the figure.
The longitudinal scaling is observed in all the models studied. 

These results then sets the stage for studying the longitudinal scaling in $v_{2}$. Note that the goal 
here is not to have a quantitative comparison with data on the scalings in $dN_{\mathrm {ch}}/d\eta$ and $v_{2}$, 
but to see if the observations are qualitatively reproduced in the models.

\section{LONGITUDINAL SCALING OF $v_{2}$}

\bef
\begin{center}
\includegraphics[scale=0.4]{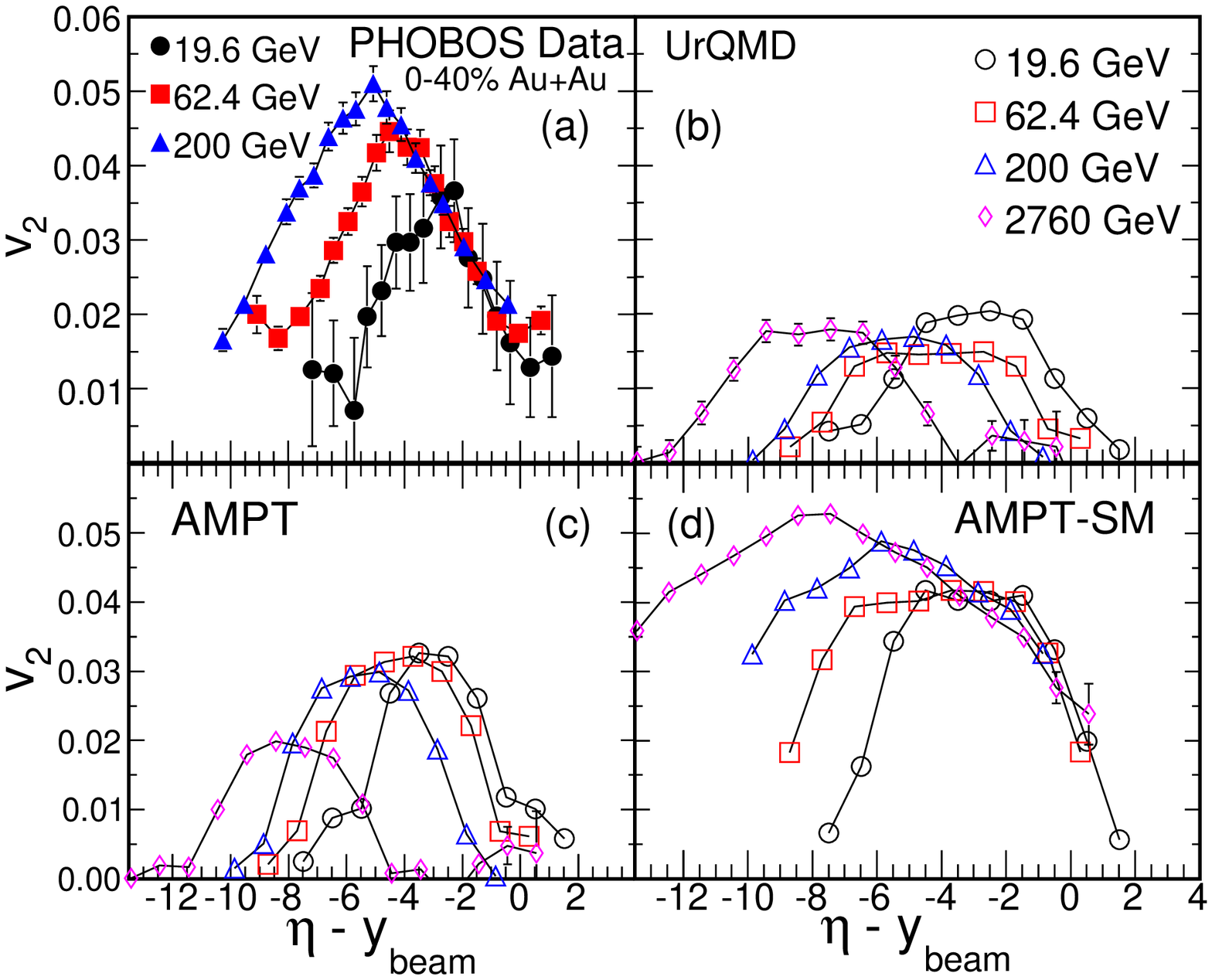}
\caption{(Color online) $v_{2}$ for charged particles versus $\eta$-$y_{\mathrm {beam}}$
for 0--40\% central Au+Au collisions at $\sqrt{s_{\mathrm {NN}}}$ = 19.6, 62.4 and 200 GeV
from (a) the  PHOBOS experiment at RHIC~\cite{v2lf}, (b) UrQMD, (c) 
AMPT default and (d) AMPT-SM. Also shown are the model results from Pb+Pb collisions 
at $\sqrt{s_{\mathrm {NN}}}$ = 2760 GeV.}
\label{fig3}
\end{center}
\eef

Figure~\ref{fig3} shows the $v_{2}$ for charged particles versus $\eta$-$y_{\mathrm {beam}}$
in Au+Au collisions at $\sqrt{s_{\mathrm {NN}}}$ = 19.6, 62.4 and 200 GeV~\cite{v2lf}. 
The results from the models for Pb+Pb collisions at $\sqrt{s_{\mathrm {NN}}}$ = 2760 GeV are also
shown. The collision centrality is 0--40\% central and is different for that shown 
for $dN_{\mathrm {ch}}/d\eta$ in Fig.~\ref{fig1}. The choice of centrality is 
based on availability of the $v_{2}$ data for charged particles in the experiment 
as a function of rapidity. 
Figure~\ref{fig3} (a) shows the longitudinal scaling of $v_{2}$ as 
measured by the PHOBOS experiment~\cite{v2lf}. Fig.~\ref{fig3} (b)  shows 
the $v_{2}$ vs. $\eta$-$y_{\mathrm {beam}}$ from UrQMD model, in (c) 
the corresponding results from AMPT default 
are shown and in (d) the same results from AMPT-SM are presented. 
It is observed that the UrQMD and the AMPT default models do not show the 
longitudinal scaling as observed in the data
(Fig.~\ref{fig3} (a)). Only the AMPT model with string melting 
qualitatively reproduces the observed longitudinal scaling of $v_{2}$. 

\section{DISCUSSIONS}
It is worthwhile to now discuss briefly the differences in these transport models.
The main difference between UrQMD and AMPT lies in the initial conditions 
(for AMPT taken from HIJING~\cite{hijing}) and additional initial state 
rescatterings in AMPT.  The main difference between AMPT default and AMPT-SM lies 
in the following: The string melting  version of the AMPT model is formulated on 
the idea that for energy densities beyond a critical value of $\sim$ 1 GeV/$fm^3$, 
it is difficult to visualize the coexistence of strings (or hadrons) and partons. 
Hence the need to melt the strings to partons. This is done by converting the 
mesons to a quark and anti-quark pair, baryons to three quarks etc. The scattering of 
the quarks are then carried out based on parton cascade~\cite{ampt}. The parton-parton
cross section taken here is 10 mb. Once the 
interactions stop, the partons then hadronizes through the mechanism of partonic 
coalescence. While for the AMPT default the scattering occurs for minijet partons 
(no melting of strings to partons) and hadronization occurs through fragmentation 
process~\cite{lund}. This model based study then suggests that partonic interactions 
in high energy density matter is essential to qualitatively reproduced  the 
simultaneous observation of the 
longitudinal scalings in $dN_{\mathrm {ch}}/d\eta$ and $v_{2}$ in experiment. 
If this is the actual cause then it will be interesting to have experimental 
measurements of $v_{2}$ vs. $\eta$ for lower beam energies where we do not expect 
to create a sufficiently high energy density system to see the breakdown of such 
a $v_{2}$ longitudinal scaling. 

We now briefly discuss some other possibilities which could explain the longitudinal scaling
of $v_{2}$. One of them is based on the arguments whether the system is weakly coupled or
strongly coupled. A weakly coupled system has been argued to favor the combined  $v_{2}$
and $dN_{\mathrm {ch}}/d\eta$ scaling behavior~\cite{torrieri}. 
It has been suggested that for systems where 
the interactions
among the constituent particles are small, or the system is close to free streaming, 
called the collisionless limit~\cite{heiselberg}, 
the  $v_{2}$ $\sim$ $\frac{dN}{d\eta}$ $\frac{\langle v \sigma \rangle}{\pi R_{x}R_{y}}$. 
Where $v$ is the relative velocity of the particles, $\sigma$ is the momentum transfer 
interaction cross section and the product $\pi R_{x}R_{y}$ is the transverse overlap area for
the two nuclei. In this model, one can easily see that  $v_{2}$ should exhibit a longitudinal
scaling similar to $dN_{\mathrm {ch}}/d\eta$ provided $\langle v \sigma \rangle$ does not 
change with beam energy. A linear dependence of $v_{2}$ with change 
in $\frac{1}{\pi R_{x}R_{y}}$$\frac{dN}{d\eta}$ has been observed in experiments over 
a wide collision systems~\cite{starflow}. In the event of $\langle v \sigma \rangle$
changing with beam energy, possibly due to change in the relevant degrees of freedom (hadronic
or partonic), there would be a breakdowm of the longitudinal scaling of $v_{2}$. This is 
consistent with the conclusions from our model study. Now let us move to the other extreme
limit, where the rescattering among the consitutent particles are abundant leading to the 
hydrodynamic limit~\cite{heiselberg}. In such a model the $v_{2}$ is proportional 
to the average transverse momentum of the particles among several other quantities as 
discussed in~\cite{flow}. If the $\langle p_{\mathrm T} \rangle$ also exhibits a 
longitudinal scaling then $v_{2}$ in the hydrodynamic limit scenario should also exhibit 
the scaling. Measuring $\langle p_{\mathrm T} \rangle$ vs. $\eta$ could help address 
the cause of the longitudinal scaling of $v_{2}$. However we have seen in Fig.~\ref{fig2}
that the models based on transport approach also exhibit longitudinal scaling of $\langle p_{\mathrm T} \rangle$.
The model study for all the three observables indicates that observing longitudinal scaling in 
 $dN_{\mathrm {ch}}/d\eta$ and/or $\langle p_{\mathrm T} \rangle$ does not neccessarily implies 
we should see a similar scaling in  $v_{2}$.

\section{SUMMARY}
In summary, we have discussed the simultaneous observation of longitudinal scaling
of $dN_{\mathrm {ch}}/d\eta$ and $v_{2}$ when plotted as a function of 
$\eta$-$y_{\mathrm {beam}}$ in RHIC experiments in Au+Au collisions at 
$\sqrt{s_{\mathrm {NN}}}$ = 19.6, 62.4, and 200 GeV. There are several
physical arguments based on convervation of entropy and hadron distributions in rapidity
corresponding to the rapidity distributions of partons (assuming hadron interactions are 
short ranged in  rapidity) to explain the longitudinal scaling
of $dN_{\mathrm {ch}}/d\eta$. The UrQMD and AMPT models qualitatively reproduce the
the experimental observations for the $\sqrt{s_{\mathrm {NN}}}$ = 19.6 - 2760 GeV. 
These models also exhibit longitudinal scaling in 
$\langle p_{\mathrm T} \rangle$. The observation of longitudinal scaling for $v_{2}$
which could be senstive to several quanities like initial conditions/densities,
equation of state and rescatterings among the consituents is bit intriguing. 
We find that among the models studied, only AMPT with string melting qualitatively
shows the behaviour as exhibited by the data. The main difference between this version
of the AMPT model compared to the default version and UrQMD lies in the partonic 
interactions and hadronization through coalescence mechanism. 
We also discussed that the longitudinal scaling can naturally occur 
for a system in the collisionless limit. In such a limit $v_{2}$ proportional only
to $dN_{\mathrm {ch}}/d\eta$ provided the product of interaction cross section and average
relative velocity of the particles does not change with the beam energy studied. 
It is expected that the interaction cross sections could be very different if the relevant
degrees of freedom are partonic or hadronic. So studying the logitudinal scaling
of  $v_{2}$ at lower and higher beam energies will provide additional insight. 
For the other extreme, hydrodynamic limit, since $v_{2}$ is proprotional 
to $\langle p_{\mathrm T} \rangle$, it would be interesting to measure in experiments 
the $\langle p_{\mathrm T} \rangle$ vs. $\eta$ for various beam energies. 
However the model study of the obsrvables $dN_{\mathrm {ch}}/d\eta$, $\langle p_{\mathrm T} \rangle$
and  $v_{2}$ indicates that observing longitudinal scaling in  $dN_{\mathrm {ch}}/d\eta$ 
and/or $\langle p_{\mathrm T} \rangle$ does not neccessarily implies that such a scaling 
will follow in  $v_{2}$.

\noindent{\bf Acknowledgments}\\
We thank Dr. G. Torrieri for discussing the topic of longitudinal scaling of 
$v_{2}$ and his helpful suggestions. X. F. Luo for help in model data at LHC energy.
Financial assistance from the Department of Atomic Energy, Government of India is 
gratefully acknowledged. PKN is grateful to the Board 
of Research on Nuclear Science and Department of Atomic Energy,
Government of India for financial support in the form of Dr. K.S. Krishnan
fellowship. LK is supported by DOE grant DE-FG02-89ER40531. BM is supported by
the DAE-BRNS project sanction No. 2010/21/15-BRNS/2026.\\

\end{document}